\documentclass[twocolumn,preprintnumbers,secnumarabic,amsmath,amssymb,superscriptaddress,nofootinbib,floatfix]{revtex4}
\usepackage{varioref,exscale,latexsym,amsmath,amssymb}

\usepackage{graphicx}
\usepackage{dcolumn}
\usepackage{bm}

\usepackage{epsfig}
\usepackage{amsmath}
\usepackage{verbatim}
\usepackage{psfrag}
\usepackage{bm}
\usepackage{dsfont}
\usepackage[footnotesize]{caption}

\newcommand{\MP}{M_{\text{P}}}

\newcommand{\ophi}{\phi^*}
\newcommand{\oW}{W^*}

\newcommand{\del}{\partial}

\newcommand{\euler}{\text{e}}

\newcommand{\be}{\begin{equation}}
\newcommand{\ee}{\end{equation}}
\newcommand{\bea}{\begin{eqnarray}}
\newcommand{\eea}{\end{eqnarray}}

\begin{document}

\preprint{CAFPE-118/09}
\preprint{MPP-2009/55}
\preprint{UG-FT-248/09}

\title{Chaotic Inflation in Supergravity with Heisenberg Symmetry}

\medskip\
\author{Stefan~Antusch}%
\affiliation{Max-Planck-Institut f\"ur Physik (Werner-Heisenberg-Institut),\\
F\"ohringer Ring 6,
80805 M\"unchen, Germany}
\author{Mar~Bastero-Gil}%
\affiliation{Departamento de Fisica Teorica y del Cosmos and\\
Centro Andaluz de Fisica de Particulas Elementales,\\
Universidad de Granada, 19071 Granada, Spain}
\author{Koushik Dutta}
\affiliation{Max-Planck-Institut f\"ur Physik (Werner-Heisenberg-Institut),\\
F\"ohringer Ring 6,
80805 M\"unchen, Germany}
\author{Steve~F.~King}%
\affiliation{School of Physics and Astronomy, University of Southampton,\\
Southampton, SO17 1BJ, United Kingdom}
\author{Philipp M. Kostka}
\affiliation{Max-Planck-Institut f\"ur Physik (Werner-Heisenberg-Institut),\\
F\"ohringer Ring 6,
80805 M\"unchen, Germany}

\begin{abstract}
We propose the introduction of a Heisenberg symmetry of the
K\"ahler potential to solve the problems with chaotic inflation in
supergravity, as a viable alternative to the use of shift
symmetry. The slope of the inflaton potential emerges from a small
Heisenberg symmetry breaking term in the superpotential. The
modulus field of the Heisenberg symmetry is stabilized and made
heavy with the help of the large vacuum energy density during
inflation. The observable predictions are indistinguishable from
those of typical chaotic inflation models, however the form of the inflationary superpotential considered here may be interpreted in terms
of sneutrino inflation arising from certain classes of string
theory.
\end{abstract}

\maketitle

\vskip 3mm

\section{Introduction}

The paradigm of inflation at the beginning of the evolution of the
Universe is very successful in explaining why the observable
Universe is homogeneous and isotropic on the largest scales, yet
with the structures on small scales \cite{Liddle:2000cg}. Among
several models of inflation, chaotic inflation is very attractive,
mainly due to its simplistic form of the potential and the
associated dynamics of the inflaton field \cite{Linde:1983gd}. On
the other hand, successful realization of chaotic inflation
requires field values which are super-Planckian. From the
observational point of view, chaotic inflation is
distinguished from small field inflation models
by its prediction of larger tensor fluctuations
\cite{Lyth:1996im}.

A generic problem of inflation models, when embedded in
supergravity (SUGRA) is that typically a large
mass for the inflaton field of the order of the Hubble scale during
inflation (which implies a slow-roll parameter $\eta \sim 1$)
is generated and spoils inflation. This
is called the $\eta$-problem in SUGRA inflation \cite{Copeland:1994vg}.
The problem arises from the SUGRA potential which
generally contains a term of the form $ e^{(K/M_{P}^2)} V_{0}$,
where $V_0$ is the vacuum energy during inflation (up to a constant).
For example, assuming a minimal canonical K\"ahler potential $K$, this becomes
$e^{(\phi \phi^{*}/ M_{P}^2)} V_{0}$, which, when expanded
in powers of $\phi \phi^{*}$, where $\phi$ is the inflaton field,
leads to a large mass term of order the Hubble constant. The problem is particularly acute in
chaotic inflation in which the inflaton field value exceeds the Planck mass
$M_{P}$.

The above problem of chaotic inflation
can be alleviated by assuming some particular non-minimal
form of the K\"ahler potential, however many of the proposed forms are not enforced
by any symmetry \cite{Goncharov:1983mw} and require some amount of tuning of parameters.
A more promising approach is to impose
a symmetry on the K\"ahler function such that the inflaton field does not explicitly
appear in the K\"ahler potential, thereby finessing
the problem of not allowing super-Planckian field values arising from
the exponential in the potential. For example, imposition of a shift
symmetry makes the K\"ahler potential independent of the real
component of the complex chiral superfield $\Phi$, thus allowing this
direction to be identified as the inflaton field \cite{Kawasaki:2000yn}.

In a recent paper \cite{Antusch:2008pn} we proposed
a class of models in which the $\eta$-problem of
supersymmetric hybrid inflation may be resolved using a Heisenberg
symmetry. For example, Heisenberg symmetry includes the case of no-scale supergravity
K\"ahler potential, which is endemic to string theory.
Due to the nature of the symmetry, the inflaton $\Phi$ appears
in the K\"ahler potential only as a combination of $\rho = T + T^* -
|\Phi|^2$, where $T$ is a modulus field \cite{Binetruy:1987xj,Gaillard:1995az}.
In this approach the associated modulus field is stabilized and made
heavy with the help of the large vacuum energy during inflation
without any fine-tuning. Because of the Heisenberg symmetry of the
K\"ahler potential, the tree-level potential of the inflaton is
flat and only lifted by radiative corrections, induced by
Heisenberg symmetry breaking superpotential couplings.

In this letter we propose the introduction of a Heisenberg symmetry of
the K\"ahler potential to solve the problems with chaotic inflation in
SUGRA. As inflationary part of the superpotential we will consider $W= M \Phi X $,
where $M$ is a mass parameter,
$\Phi$ is the superfield containing the inflaton and $X$ is an auxiliary superfield.
\footnote{We require the chaotic superpotential to be of this form,
rather than the more common form $W= M \Phi^2$ since, according to our mechanism,
the field $X$ will quickly settle to a zero value during inflation,
resulting in the superpotential also being zero during inflation, which is a crucial
requirement for solving the $\eta$ problem in this approach.}
This form of the superpotential has been
discussed for chaotic inflation in \cite{Kawasaki:2000yn}, where a shift symmetry has been used to protect
the flatness of the inflaton potential.
Using the Heisenberg symmetry with this form of the superpotential,
we show that the $\eta$-problem of SUGRA inflation is resolved, allowing well behaved inflaton
field values larger than the Planck mass, as required for a successful realization
of chaotic inflation. We also show that,
as in the case of hybrid inflation \cite{Antusch:2008pn},
the associated modulus field will be stabilized during
inflation by the large vacuum energy. It turns out that the dynamics of the model will be
effectively governed by a single scalar field with a tree-level potential
which has a purely quadratic dependence on the inflaton field. (We estimate
the effects of quantum corrections and show that they are negligible.) Thus we find that the
observable predictions for the minimal setup discussed in this paper
are indistinguishable from those of typical chaotic inflation models.
However, the framework proposed here will have
important implications when the theory of inflation is
embedded into a particle physics model.

The paper is organized as follows: In the next section we present
the framework of our model. In section 3, we discuss the effective
tree-level scalar potential. The dynamics of all the scalar fields is
discussed in detail in section 4. In section 5 we summarize and conclude.
In the Appendix we show that the quantum loop corrections
do not change the tree-level potential.

\section{Framework}\label{The Model}

To implement chaotic inflation in SUGRA using Heisenberg symmetry we
propose the following superpotential and K\"ahler potential,
\begin{eqnarray}
\label{superpotential} W&=& M \Phi X \:, \\
\label{Kaehlerpotential} K &\equiv& (1 + \kappa_X |X|^2 + \kappa_\rho \rho) |X|^2 +
f(\rho) \:.
\end{eqnarray}
In the K\"ahler potential the modulus $T$ together with the
chiral inflaton superfield $\Phi$ appear only in the Heisenberg symmetric
combination \cite{Binetruy:1987xj}
\be \rho = T + T^* - |\Phi|^2 \,.
\ee
Here $X$ is an additional superfield that is supposed to be fixed at zero
when inflation is operative.  The non-zero F-term from $X$
contributes to the vacuum energy required for inflation.  Therefore, we
realize inflation in this framework with vanishing inflationary part of the
superpotential.  We note that the superpotential of the form of
Eq.~\eqref{superpotential} breaks the Heisenberg symmetry of the
K\"ahler potential in Eq.~\eqref{Kaehlerpotential} and gives rise to
a $M^2 \phi^2$ potential for the scalar component of $\Phi$ at tree-level.
This model is `natural' in the sense that setting the small breaking parameter to
zero allows us to realize enhanced symmetry \cite{'tHooft:1980xb}. We
take all chiral superfields to be gauge singlets, such that D-term
contributions to the potential are absent.

\section{Scalar Potential}\label{potential}

A Heisenberg symmetry of the K\"ahler potential allows us to implement
super-Planckian values for the inflaton field in SUGRA theories, as
can be seen by looking at the full (F-term) scalar potential given by
\footnote{Here we use a convention in which we set the reduced Planck
mass $\MP\simeq 2.4\times 10^{18}\,\text{GeV}$ to unity.}
\be
\label{Fterm}
V_{\text{F}}=\text{e}^K\left[K^{i\bar{j}}\,
\mathcal{D}_{i}W\,\mathcal{D}_{\bar{j}}\oW - 3|W|^2\right]\,,
\ee
where the definition
\be
\mathcal{D}_{i}W:=W_{i}+K_{i}\,W
\ee
has been used.
The lower indices $i,j$ on the superpotential or K\"ahler potential denote the
derivatives with respect to the chiral superfields or their conjugate
(where a bar is involved). The inverse K\"ahler metric is defined as
$K^{i \bar j} = K^{-1}_{i\bar j}$.  Due to the Heisenberg symmetry of
the K\"ahler potential the exponential in Eq.~\eqref{Fterm} is
independent of the inflaton field and therefore we can realize field
values larger than $M_{P}$ as required for chaotic
inflation.

We will show below that the mass of the $X$ field is very large compared to
the Hubble scale due to the $\kappa_X$ coupling in the K\"ahler
potential and it settles to its minimum at $X=0$ very quickly before
inflation. This essentially makes the superpotential $W$ of Eq.~\eqref{superpotential} vanish during
inflation. An attractive feature of having $W=0$ during inflation is that
it typically cancels several couplings between the inflaton sector and
any other possibly existing scalar field sector in the theory
\footnote{In the context of Hybrid inflation with shift symmetry, see
Ref.~\cite{Antusch:2009ef}.}.

The K\"ahler metric can be calculated as the second derivatives of the
K\"ahler potential in Eq.~\eqref{Kaehlerpotential} with respect to the
superfields and their conjugates, and along the direction $X=0$ in
$(X,\Phi,T)$-basis
this reduces to the block-diagonal form
\be
\label{KaehlermetricInflation}
 \left(K_{i\bar{j}}\right)=
 \begin{pmatrix}
1+\kappa_{\rho}\,\rho & 0 & 0\\ 0 & f''(\rho)|\phi|^2 - f'(\rho) &
-f''(\rho)\ophi \\ 0 & -f''(\rho)\,\phi & f''(\rho)
 \end{pmatrix}\,.
\ee
Here $\phi$ denotes the complex scalar component of the inflaton superfield
$\Phi$. For $X=0$ the potential reduces to
\be
V_{\text{F}} = e^{K}\,K^{X \bar X}\,|W_X|^2\,.
\ee

Now we make a particular choice of $f(\rho)$ of the no-scale form
\cite{noscale1}
\be\label{noscaleeq}
f(\rho)= -3 \ln \rho \,.
\ee
It provides a simple example of a K\"ahler potential invariant under
the Heisenberg symmetry and arises naturally
in orbifold compactifications of heterotic string models
\cite{noscale2}.
We would like to remark that for our approach to work this specific form of the
K\"ahler potential is not required, however it is well motivated from string theory and
serves well to illustrate the modulus stabilization mechanism through the
coupling with $X$ in the K\"ahler potential.
After inflation, when $X \sim \Phi \sim 0$, another sector of the model will
be responsible for SUSY breaking which may lead, for instance, to an
effective no-scale model with radiatively induced gravitino mass.

With $f(\rho)$ as in Eq.~(\ref{noscaleeq}), the potential reduces to the simple form
\be\label{treepotential}
V_{\text{F}}= \frac{M^2|\phi|^2}{\rho^3 ( 1 + \kappa_\rho \rho)} \,.
\ee
For any fixed value of $\rho$ the potential is just a mass term
for the $\phi$ field that is suitable for chaotic inflation.
This potential has a minimum at
\footnote{For a general function $f(\rho)$, the minimum is given by\\
  $f^\prime(\rho_0) (1 + \kappa_\rho \rho_0)=\kappa_\rho$.}
\be \rho_{0}=
-\frac{3}{4 \kappa_\rho}\,.
\ee
Considering that $\rho$ is always positive, $\kappa_{\rho}$ must be
negative and, as an example, we will choose its value to be
$\kappa_{\rho}=-1$. Other than that, the form
of the potential has a pole at the value of $\rho = -1/\kappa_{\rho}$
and for values $\rho > -1/ \kappa_{\rho}$ the potential is negative
and has a runaway behavior. Therefore, inflation happens only in the
range of field values $ 0< \rho < -1/\kappa_{\rho}$ and we assume it
for all our considerations.

At the start of inflation $m_{\rho}^2 \sim |W_{X}|^2$ and the squared Hubble
scale is also of the same order.  On the other hand $m_{\phi} \sim
M \ll H$ during inflation.  Therefore the $\rho$ field settles to its
minimum very quickly whereas the inflaton field, being light, slow-rolls
along its potential.  When $\rho$ has settled to its minimum and $\phi$ is
slowly rolling the vacuum
energy dominates and drives inflation.  In fact, the coupling
$\kappa_\rho$ between $\rho$ and $X$ in the K\"ahler potential induces
a mass for the $\rho$ field of the order of the vacuum energy
during inflation and it allows the modulus to be stabilized very
quickly before inflation.
For $\kappa_{\rho}=0$ in the expression for the scalar
potential of Eq.~\eqref{treepotential}, we can see that the $\rho$ field would have a runaway potential.
The role of this coupling in the context of hybrid inflation with
Heisenberg symmetry was first mentioned in Ref.~\cite{Antusch:2008pn}.
Although $\rho_{0}$ is independent of $\phi$, the field $\rho$ is not
absolutely fixed at the minimum of the potential due to the presence
of effects from non-canonical kinetic terms. We will verify these qualitative
statements in the next section with full evolution equations of the
fields.  We also note that when $X$ and $\rho$ settle to their
respective minima (i.e. $X = 0$ and $\rho = \rho_{0}$) the potential
has the same form as the corresponding global SUSY potential.
For example in our case, once the modulus is stabilized at the minimum during inflation, the
tree-level potential reduces to a quadratic potential,
as can be seen from Eq.~\eqref{treepotential}.
We will discuss the scalar mass spectrum in the following section.

\section{Dynamics of the fields}\label{dynamics}

In the previous section we have discussed the form of the potential when
the $X$ field has settled to its minimum and argued qualitatively
that the $\rho$ field will also get stabilized quickly such that successful
chaotic inflation can be realized. In this section we calculate the
masses of the fields and perform a full numerical simulation to investigate the
dynamics of the fields.

We begin with writing down the kinetic energy terms for the fields.
With $X=X^*=0$ in the K\"ahler metric, the kinetic terms are given by:
\be {\cal L}_{\text{kin}} = (1 + \kappa_\rho \rho) \,| \partial_\mu
X|^2 + \frac{3}{\rho}\,|\partial_\mu \phi |^2 + \frac{3}{4
\rho^2}\,(\partial_\mu \rho)^2 + \frac{3}{4 \rho^2}\, (I_\mu)^2 \,,
\label{Lkin}
\ee
where
\be I_\mu= \text{i}\, (\partial_\mu(T -T^*) +
\phi\,\partial_\mu \phi^*- \phi^* \,\partial_\mu \phi) \,.
\ee
Since the phases of the scalar fields $\phi$, $X$ as well as $\mbox{Im}(T)$ very quickly
approach a constant value in an expanding Universe and subsequently decouple from
the absolute values and $\mbox{Re}(T)$ in the equations of motion (as has been discussed in the
Appendix of \cite{Antusch:2008pn}), we only consider the absolute values $\varphi = \sqrt{2}\,| \phi |$,
$x =\sqrt{2}\, |X|$ and the real field $\rho$ in what follows.
In this case, $I_{\mu}$ vanishes identically.

The evolution equations for the background field values $\rho$,
$\varphi$ and $x$ in an expanding universe are then given by:
\be
\begin{split}
\ddot x +3\,H\,\dot
x+\frac{1}{(1+\kappa_{\rho}\rho)}\,\left(\kappa_{\rho}\,\dot\rho\,\dot
x+V_{x}\right)&=0\,,\\ \ddot \varphi + 3 \,H\, \dot \varphi
-\frac{1}{\rho}\,\dot \rho\, \dot \varphi + \frac{\rho}{3}\,
V_{\varphi} &=0 \,,\\ \ddot \rho + 3 \,H \,\dot \rho- \frac{1}{\rho}
\,\dot \rho^2 + \dot \varphi^2 -\frac{\kappa_{\rho}}{3}\,\rho^2\,\dot
x+ \frac{2}{3}\, \rho^2 \,V_\rho&=0 \,,
\end{split}
\ee
where $V_\rho$, $V_{\varphi}$, $V_{x}$ are the derivatives of the
potential with respect to the fields, and $H$ is the Hubble expansion
rate.  For large enough values of $\varphi$, the fields follow a
slow-roll trajectory with $\rho$
being practically at the minimum of the potential.

\begin{figure}[htbp!]

\psfrag{MP}{$x,\varphi,\rho\,[M_{\text{P}}]$} \psfrag{Mt}{$M\,t$}
\includegraphics[width=0.45\textwidth]{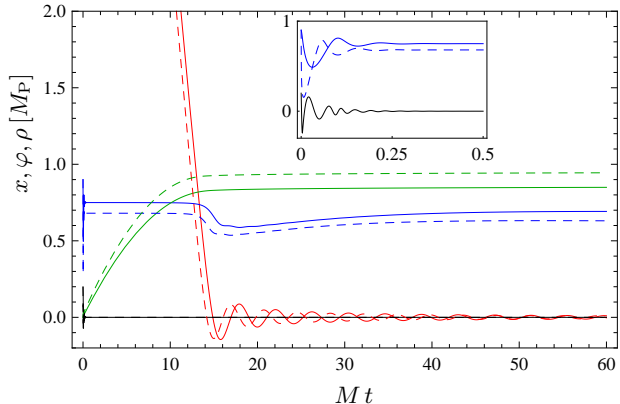}

\caption{Evolution of the fields. The green line represents the number of
e-folds/100, the red line represents the inflaton direction $\varphi$,
the blue line depicts the evolution of $\rho$ and the black line the
evolution of $x$.  Dashed lines represent the evolution of the same fields if in
addition to the tree-level potential, quantum corrections are
taken into account (see Appendix).}
\label{evol.eps}
\end{figure}

In Fig.~\ref{evol.eps} we show the dynamics of the fields for generic
initial conditions.  As one can see, the $x$ field settles to its
minimum followed by the $\rho$ field, whereas the inflaton field
remains slowly rolling for more than 60 e-folds of inflation. As the inflaton
field rolls down towards its minimum, the vacuum energy decreases and
thus also the mass of the $\rho$ field. When the slow-roll conditions
are violated the inflaton field acquires a large velocity and it
provides a kick to the evolution of $\rho$ due to the $\dot
\varphi^2$ term in its equation of motion.  At the end of inflation
the $\varphi$ field starts oscillating, the velocity term gets damped
and finally the $\rho$ field settles to a slightly different field
value. However, at this epoch we expect some other
moduli-stabilization mechanism to start playing a role.

In the following, we denote as physical fermions and scalars those with canonical kinetic
terms. In terms of the original fields $\Psi_i = \{ X,\phi,T \}$, the
Lagrangian for the scalar kinetic terms reads:
\be {\cal L}_{\text{kin}} = K_{ij*}\, \partial_\mu \Psi_i \,\partial^\mu
\Psi^*_j \,,
\ee
and similarly for the fermions. We can canonically normalize the
fields by expanding the Lagrangian around the
minimum.  In particular, during inflation we have $X=0$, $\rho\simeq
\rho_{0}=-3/4\kappa_\rho$, and $\phi$ changing slowly with time.  The
field $\rho$ can be redefined as canonically normalized field $\tilde \rho$
by:
\be
\tilde \rho =\sqrt{\frac{3}{2}} \ln \rho \,.
\ee
The remaining non-canonical factors in Eq.~\eqref{Lkin} only depend on
$\rho$ (or $\tilde \rho$), and we can define the physical states
during inflation by expanding those factors around
$\rho_{0}$:
\be {\cal L}_{\text{kin}} = (1 + \kappa_\rho \rho_0)\, |
\partial_\mu X|^2 +\frac{3}{\rho_0}\, |\partial_\mu \phi |^2 +
\frac{1}{2} \,(\partial_\mu \tilde \rho)^2 + \cdots\,.
\ee
Then, with normalization factors $(1 + \kappa_\rho \rho_{0}\,,\, 3/\rho_{0}\,,\, 1)$,
for ($\tilde x$, $\tilde \varphi$, $\tilde \rho$), the physical squared scalar masses are given by:
\bea
m^2_{\tilde x} &=& \frac{64}{27}\,\kappa^2_{\rho}\, M^2 \,(1 +32 \,\kappa_X\,\kappa_{\rho}\, \tilde
\varphi^{2})\:,\\
m^2_{\tilde \varphi}&=& -\,\frac{64 }{27}\,\kappa^3_{\rho}\, M^2
\label{inflatonmass}\:,\\
m^2_{\tilde \rho} &=& -\,\frac{256}{27} \,\kappa^3_{\rho}\, M^2\, \tilde\varphi^2\:.
\eea
Fields with a tilde are canonically normalized. To calculate their masses, we have assumed that $\rho$ has settled to its minimum (where $\kappa_\rho$ has been left general instead of setting it to $-1$ as done before).
First we note that the
canonically normalized inflaton field has a mass smaller than the
Hubble scale during inflation where the field value is
super-Planckian. For $\kappa_X$ negative, the mass
of the $\tilde x$ field is larger than the mass of $\tilde \rho$ and
both masses are larger than the Hubble scale $H \sim \sqrt{V/3}$.
Therefore, as we have shown in the numerical simulations, the $x$
field settles quickly to its minimum $x = 0$, followed by the $\rho$ field.

Similar to the conventional chaotic inflation model with purely quadratic potential the prediction for the
spectral index $n_s$ and the tensor-to-scalar ratio $r$ are given by
\begin{eqnarray}
n_s &\simeq& 1 - 2/N\:, \\
r   &\simeq& 8/N \:,
\end{eqnarray}
where $N$ is the number of e-folds before the end of inflation where the observable scales have crossed the horizon. For $60$ e-folds of inflation the predicted value
of the spectral index is $n_s \sim 0.97$, well consistent with all
available cosmological data \cite{Komatsu:2008hk}. In addition,
the tensor-to-scalar ratio is predicted as $r \sim 0.13$ (for $N=60$),
which might be probed by the upcoming PLANCK satellite \cite{planck}.

\section{Conclusions}\label{conclusions}

We have proposed the introduction of a Heisenberg symmetry of
the K\"ahler potential to solve the problems with chaotic inflation in
SUGRA, as a viable alternative to the use of shift symmetry.
We have focussed on a particular form of chaotic inflation based on
the superpotential $W= M \Phi X $, where $M$ is a Heisenberg symmetry breaking mass parameter,
$\Phi $ is the superfield which contains the inflaton and $X$ is an auxiliary superfield.
Within this framework we have shown that the $\eta$\,-\,problem
of the SUGRA potential is resolved, allowing well behaved inflaton
field values larger than the Planck mass, as required for a successful realization
of chaotic inflation. We have also shown that
the associated modulus field is stabilized during
inflation by the large vacuum energy.

We have seen that the dynamics of the model is
governed by a single scalar field with a tree-level potential
which has a purely quadratic dependence on the inflaton field.
In an Appendix we have estimated
the effects of quantum corrections and shown that they are negligible.
Thus we find that the
observable predictions for the minimal setup discussed in this paper
are indistinguishable from those of typical chaotic inflation models,
namely a spectral index $n_s \sim 0.97$ and a tensor-to-scalar ratio $r \sim 0.13$.

However the framework proposed here will have
important implications when the theory of inflation is
embedded into a particle physics model. For example, the
effective mass of the inflaton differs
from the mass scale $M$ by a factor which depends on
$\kappa_{\rho}$ (cf. Eq.~\eqref{inflatonmass}). If the present chaotic inflation model
is interpreted as a (right-handed) sneutrino inflation model \cite{Murayama:1992ua},
then this would change the mass scale of the associated right-handed
neutrinos. It is also worth remarking that the type of chaotic superpotential
considered here would correspond to Dirac right-handed neutrino
masses of the form $MN_1N_2$ where $N_1$ is identified with $\Phi$
and $N_2$ is identified with $X$. Such a Dirac right-handed neutrino
mass structure is motivated by certain classes of Heterotic string theory
\cite{Giedt:2005vx}. It would be interesting to discuss such a Dirac sneutrino
interpretation of the inflation model considered here, and the associated
implications for leptogenesis, however this would
require an analysis of the evolution of the $\rho$ field after inflation
which is beyond the scope of this letter.
The embedding into a particle physics model would also allow
to address the issue of reheating.
For instance, in sneutrino inflation, the inflaton would decay 
via its neutrino Yukawa couplings.

In conclusion, we have proposed the introduction of a Heisenberg symmetry of
the K\"ahler potential to solve the problems with chaotic inflation in
SUGRA, as a viable alternative to the use of shift symmetry. It leads to the
same predictions for $n_s$ and $r$, however has different implications when embedded
into a particle physics model.

\section*{Appendix: Calculations of Quantum Loop Corrections}

Here we discuss that the quantum corrections to the potential have
negligible effects.  We therefore calculate the one-loop effective
potential from the formulae given in~\cite{Gaillard:1995az, Ferrara:1994kg}.
Introducing a cutoff $\Lambda=\MP$ in the theory, the one-loop
correction to the effective potential is given by
\be
V_{\text{loop}}=\frac{\Lambda^2}{32 \pi^2}\,\text{Str}\, \mathcal{M}^2
+\frac{1}{64\pi^2}\,\text{Str}\,\mathcal{M}^4\,
\ln\left(\frac{\mathcal{M}^2}{\Lambda^2}\right)\,.
\ee
Since in order to fit observations, the mass parameter $M$ should be
of the order $M\sim\mathcal{O}(10^{-5})$, we can safely ignore the
logarithmic part of the loop-correction.

For the dominant quadratic part, the supertrace can be written in the
simple form
\be
\label{supertrace}
\begin{split}
\text{Str}\,\mathcal{M}^2=& \,\,2\, (N_{\text{tot}}-1)\,V \\
&+ 2\,|W|^2\,\euler^K
\left(N_{\text{tot}}-1-G^i\,R_{i\bar{j}}\,G^{\bar{j}}\right)\,,
\end{split}
\ee
where $N_{\text{tot}}=3$ is the total number of chiral superfields
and the K\"ahler function reads
\be
G=K+\ln\,|W|^2\,.
\ee
The contribution due to the Ricci tensor of the K\"ahler manifold in
Eq.~\eqref{supertrace} is the following
\be
R_{i\bar{j}}=\del_{i}\,\del_{\bar{j}}\,\ln\,\det\,(G_{m\bar{n}})\,.
\ee

Taking $\kappa_X <0$, the curvature along the $X$-direction is large
and positive during inflation when $\varphi \ne 0$, so that the field
will quickly go to zero.  For the sake of simplicity, we set $x=0$
which is justified by the simulation with the full $x$-dependent
potential as depicted in Fig.~\ref{evol.eps}.  Plugging the
superpotential and K\"ahler potential
of Eqs.~\eqref{superpotential},~\eqref{Kaehlerpotential} into
Eq.~\eqref{supertrace}, we end up with the $\rho$- and
$\varphi$-dependent loop-potential
\be
V_{\text{loop}}=\frac{M^2 \varphi^2}{32\,\pi^2\,(3\,\rho^3)}\, \left[\frac{2 \,(3 + 4 \kappa_{\rho} \,
\rho)}{(1+\kappa_{\rho}\, \rho)^2} -\frac{( \kappa_{\rho} \rho+12\,
\kappa_{X})}{(1+\kappa_{\rho}\, \rho)^3}\right] \,.
\ee
Therefore, the presence of loop-corrections just has the effect of
shifting the minimum of the potential in the $\rho$-direction
(cf. Fig.~\ref{evol.eps}). For example with $\kappa_{\rho} = -1$ and
$\kappa_X = -1$, the loop-corrected minimum shifts to $\rho_{0} = 0.68$ from
its tree-level value of $\rho_{0} = 0.75$. It is important to
note that similar to the tree-level potential, the loop-corrected
potential also has a $\varphi$-independent minimum $\rho_{0}$. Once the
$\rho$-field
gets stabilized in its new minimum, this gives a different factor in
front of the mass squared.  In order to fit the amplitude of the
curvature perturbation $P_{\mathcal{R}}$, one simply absorbs this
factor in $M^2$ and adjusts the new effective mass squared.
The amplitude of the primordial spectrum is given by $P_{\mathcal{R}}^{1/2}
\simeq (N / \sqrt{6}\, \pi) \,m_{\text{eff}}$, and the WMAP normalization by
$P_{\mathcal{R}}^{1/2} \simeq 5 \times 10^{-5}$ then gives $m_{\text{eff}}
\simeq 6\times 10^{-6}$. We therefore conclude that the
loop-corrections do not change the predictions calculated from the
tree-level potential but instead just lead to a mass-renormalization
of the inflaton field.

\section*{Acknowledgments}
S.F.K.\ acknowledges partial support from the following grants:
STFC Rolling Grant ST/G000557/1; EU Network MRTN-CT-2004-503369; EU ILIAS RII3-CT-2004-506222.
S.A.\ , K.D.\ and P.M.K.\ were partially
supported by the the DFG cluster of excellence ``Origin and Structure
of the Universe''.  The work of M.B.G.\ is partially supported by the
M.E.C.\ grant FIS 2007-63364 and by the Junta de
Andaluc\'{\i}a group FQM 101.

\end{document}